\documentclass[aps,prl,twocolumn,superscriptaddress,showpacs]{revtex4-1}
\usepackage{fullpage}
\usepackage{mathrsfs} \usepackage{amssymb} \usepackage{graphicx,color} \usepackage{bbm} \usepackage{multirow}
\usepackage{bm} \usepackage{mathrsfs} \usepackage{commath} \usepackage{stmaryrd} \usepackage{color}

\usepackage[latin1]{inputenc}
\usepackage{ulem}
\usepackage{amsmath}
\usepackage{amssymb}
\usepackage{bm}

\newcommand{\beq}{\begin{equation}}
\newcommand{\eeq}{\end{equation}}

\begin{document}

\title{Eigenstate thermalization and rotational invariance in ergodic quantum systems
}

\author{Laura Foini}
\affiliation{IPhT, CNRS, CEA, Universit\'{e} Paris Saclay, 91191 Gif-sur-Yvette, France}

\author{Jorge Kurchan} 
\affiliation{Laboratoire de Physique de l'ENS, Ecole Normale Sup\'erieure, PSL Research University, Universit\'e Paris Diderot, Sorbonne Paris Cit\'e, Sorbonne Universit\'es, UPMC Univ. Paris 06, CNRS, 75005 Paris, France}

\begin{abstract}
Generic rotationally invariant matrix models satisfy a simple relation: the probability distribution of half
the difference between any two diagonal elements and the one of off-diagonal elements are the same.
 In the spirit of the Eigenstate Thermalization Hypothesis (ETH),
we test the hypothesis that the same relation holds in quantum systems that are non-localized,
when one considers small energy differences. The relation provides a stringent test of ETH
beyond the Gaussian ensemble. We apply it to a disordered spin chain, the SYK model and a Floquet system.
\end{abstract}

\date{\today}

\maketitle

{\it Introduction.} ETH has become one of the most accepted elements for our understanding
of the dynamical and thermalization properties of  quantum systems~\cite{srednicki1994chaos,srednicki1999,dalessio2016}.
 It relies on the idea that the eigenvectors
of the Hamiltonian of a chaotic quantum system are affected by a  degree
of randomness and aims to quantify 
its statistical properties.
The initial assumption, based on work of Berry \cite{berry1977}, 
was that in certain simple systems like billiards, wavefunctions could be considered as a random superposition of plane waves.
Subsequent works by Deutsch and Srednicki \cite{deutsch1991,srednicki1999} extended this 
idea in order to characterize the fluctuations
of matrix elements of physical operators, from element to element and under different realizations of the Hamiltonian~\cite{dalessio2016}. 
Altogether these assumptions allow us to explain why 
locally a closed quantum system evolving from some excited state approaches equilibrium
and to estimate dynamical correlation functions~\cite{srednicki1994chaos,srednicki1999,dalessio2016}. 
ETH leads
to an ansatz for the elements of a generic local observable in the energy eigenbasis given by:
\begin{equation}
A_{ij} \sim {\cal{A}}(E) \delta_{ij} + e^{-S(E)/2} f(E,\omega) R_{ij}
\label{etheth}
\end{equation}
where $E= (e_i+e_j)/2$ and $\omega= (e_i-e_j)/2$ and $f$ and ${\cal{A}}$ are smooth functions of $E/N$ and $\omega$ and $S(E)$ is the entropy.
The $R_{ij}$ is a Hermitean matrix with essentially random elements. 
To a first approximation 
the matrix elements $R_{ij}$ are assumed to be 
Gaussian and uncorrelated random variables, but subsequent work has
shown that, strictly speaking, they are neither: 
recently \cite{foini2019}, we argued that the matrix entries $A_{ij}$ cannot
be in general independent and that the  products of certain off-diagonal matrix elements 
should have  small but relevant expectations which enter in the computation of higher order
correlation functions. This, together with the fact that the probability distributions 
of single elements are measurably non-Gaussian
\cite{luitz2017,luitz2016}, calls 
 for a better understanding of the link between ETH and random matrix theory. In the second half of this paper
 we shall assess this connection  numerically for several standard quantum models.
 
The relation between ETH (for small $\omega$) and random matrix ensembles is based upon 
the argument \cite{deutsch1991} 
according to which eigenvectors of a non-localized  Hamiltonian with near energies mix with essentially random
phases when a small perturbation is applied. 
This naturally leads  one to assume 
that matrix elements close to the diagonal (where the mixing is maximal)
are well represented by
matrices such that the joint probability density of their elements $P(A) \equiv P(\{A_{ij}\})$ is left invariant
by change of basis,  i.e. 
\begin{equation}
P(A) = P(U^\dag A U)
\label{rotational}
\end{equation}
where $U$ may be a orthogonal, unitary or symplectic matrix~\cite{mehta2004}.
A subclass of well-studied models
that enjoys this property has  elements of the form \cite{di19952d} 
$P(A) \propto \exp\left(-  \frac 1{2}N \text{Tr} V(A)\right)$
where $V(A)$ is some generic function.
The Gaussian ensemble $V(A)=A^2/2$ is the only one which 
has both rotational symmetry and independence of matrix entries \cite{livan2018}. 
 The rotational invariance  (\ref{rotational}) is a weaker condition than gaussianity, but implies testable relations between the joint distributions of 
elements \cite{Foini_prep} that, in particular, do not involve determining the distribution itself. 
These relations are easy to prove
for a random ensemble, but for a quantum system remain an assumption,  a check of the randomness of the
diagonalizing basis below the Thouless energy scale~\cite{dalessio2016}. 
In this paper we explore the simplest one, relating the distribution of diagonal and off-diagonal matrix elements.
We shall see that the validity of the relation in a quantum model is a stringent test of eigenvector thermalization, and may be used
together with the  standard study of level-spacings.

{\it Ensembles of Matrices:} 
For the purposes of computing large deviations, for example of diagonal matrix elements, we could envisage 
using the  elements of a single matrix. The problem is that there are not enough of those if we wish to compute the tails
of the distribution: we are hence forced to consider a distribution of matrices. 
Furthermore, when we will apply our conclusions to a quantum system, 
 unlike the case of random matrices, we shall not be free to rotate
 the eigenvectors as we please without destroying the structure
of the matrix, so we shall 
need to clarify from which ensemble the matrices are chosen. 
Following Deutsch \cite{deutsch1991}, we may consider the original extensive Hamiltonian modified
by an ensemble of random perturbations that are also local and have the same symmetries, but are small: sub-extensive in the system's size.
Averaging over this ensemble, we may  restrict ourselves (at least in principle) to computing the elements $A_{ii}$, $A_{jj}$ and $A_{ij}$ for any {\it fixed pair} $(i,j)$ 
with $\omega=(e_i-e_j)$ within the Thouless scale. 

{ {\it A simple relation.}} 
In order to understand what may be the relation between the probability distributions of diagonal and off-diagonal elements
of an operator, we may get inspiration in the corresponding relation for a generic rotationally invariant ensemble
as in Eq (\ref{rotational}). Let us then consider an $N\times N$ matrix  $A$ with probability density  $P(A)$, and 
focus on the (marginal) joint probability distribution $\hat P(\hat A)$
of the elements of a chosen $2 \times 2$ submatrix $\hat A$ such that
$A_{\alpha_1 \alpha_1}=\hat A_{11}$, $A_{\alpha_2 \alpha_2}=\hat A_{22}$ and $A_{\alpha_1 \alpha_2}=\hat A_{12}$
with $\alpha_1=i$ and $\alpha_2=j$.
Restricting for simplicity to the orthogonal ensemble (the generalization to unitary is straightforward) we have:
\begin{eqnarray}\label{annealed}
& &\hat P (\hat A)  = \nonumber 
 \int \mathcal{D} A \; P(A) \prod_{k\leq l=1}^2 \delta ( A_{\alpha_k \alpha_l}  - \hat A_{kl})   \nonumber \\  
\end{eqnarray}
with $\mathcal{D} A = \prod_{i=1}^N {\rm d} A_{ii} \prod_{i > j} {\rm d} A_{ij}$.
Let us now evaluate $\hat P (T^\dag \hat A T)$ where $T$ is a $2\times 2$ orthogonal matrix:
\begin{eqnarray}
&\hat P (T^\dag \hat A T) = \int \mathcal{D} A \; P(A)  \prod_{k\leq l=1}^2 \delta ( A_{\alpha_k \alpha_l}   -T_{vk} T_{ul} \hat A_{vu})     \nonumber \\  
 \vspace{-0.2cm}\nonumber \\
& = \int \mathcal{D} A \; P(A) \prod_{u\leq v=1}^2 \delta (T_{uk} T_{vl} A_{\alpha_k \alpha_l}  - \hat A_{uv})    \label{coss}
\end{eqnarray} 
where the sums over repeated indices are implicit
and the equality holds because in the delta functions we make a change of variable with Jacobian equal to one.
Let us now consider the $N\times N$ orthogonal transformation
$U_{kl} = T_{11} \delta_{ki} \delta_{li}+ T_{22} \delta_{kj} \delta_{lj} + T_{12} \delta_{ki} \delta_{lj} + T_{21} \delta_{kj} \delta_{li} 
+ \delta_{kl}(1-\delta_{ki})(1-\delta_{kj})$.
In the integral (\ref{coss}) we can make the change of variable
$A'=UAU^{\dag}$ and exploiting the rotational invariance
of the overall probability distribution $P(A)$ and of the measure
$ \mathcal{D} A$ we see that
\begin{equation}\label{Eq_rot_inv_P}
\hat P(\hat A) = \hat P(T^{\dag} \hat A T)
\end{equation}
i.e. the submatrix inherits the symmetries of the original matrix.
This invariance implies that the joint probability distribution $\hat P (\hat A)$ of all elements of the submatrix $\hat A $ depend only on its eigenvalues $\tilde A_{a}$.

For a two by two submatrix (with any $(i,j)$),
in terms of 
the difference $\hat A_-=(\hat A_{ii}-\hat A_{jj})/2$, the sum $\hat A_+= (\hat A_{ii}+\hat A_{jj})/2$ and the off-diagonal element $\hat A_{ij}$, we have:
\begin{equation}
P(\hat A) =F\left(  \hat A_+ + \sqrt{\hat A_-^2+ \hat A_{ij}^2} , \hat A_+ - \sqrt{\hat A_-^2+ \hat A_{ij}^2} \right) 
\label{eigen1}
\end{equation}
This function is symmetric with respect to exchange of $\hat A_-$ and $\hat A_{ij}$. 
This means that their probability marginals 
are the same function $  \bar F$ evaluated in $\hat A_-$ and $\hat A_{ij}$, respectively,
and hence we conclude that they are equally distributed:
$P(\hat A_{ij}) = \bar F(\hat A_{ij})$  and $P(\hat A_{-}) = \bar F(\hat A_{-}) $.
The same may be said of the real and imaginary parts of the off-diagonal elements in the complex case.

{\it Large $N$ results.}
The previous relation holds -- in a matrix model with rotational invariance  --  for any system
size.
For a distribution $P(A) \propto \exp\left(-  \frac{1}{2}N \text{Tr} V(A)\right)$  one can determine the probability distribution of 
matrix elements (or their real and imaginary part for 
hermitian operators) in the large $N$ limit, in a general way.
They take the form of a  large deviation function,
for instance for the diagonal matrix elements  \cite{GM_05,Foini_prep}:
\beq\label{Large_dev}
P_d(A_{ii}) \simeq  e^{- \frac{N }{2} \text{extr}_s \left(  s A_{ii} - \int_0^s R(x) {\rm d} x \right)}
\eeq
where $R$ is a function dependent on the potential $V$, the  `$R$-transform'  of the spectrum \cite{tulino2004}.
The distribution of off-diagonal matrix elements 
can be written at the saddle point level:
\beq\label{Eq_conv}
P_o(A_{ij}) 
\simeq 2  \int {\rm d} A  P_d(A) P_d(A - 2 A_{ij})
\eeq
which is the large $N$ limit of our result  when the probability of  $A_{ii}$ and $A_{jj}$ factorize.

The exponential scaling (\ref{Large_dev}) suggests that the probability distributions of 
matrix entries of quantum operators are also large
deviation functions (with some `effective $N$' related to level-spacing)  and 
the deviations from the Gaussian behavior should be visible only in the tails
of those distributions for small system sizes.

\begin{figure}
\centering \includegraphics[angle=0,width=0.4\textwidth]{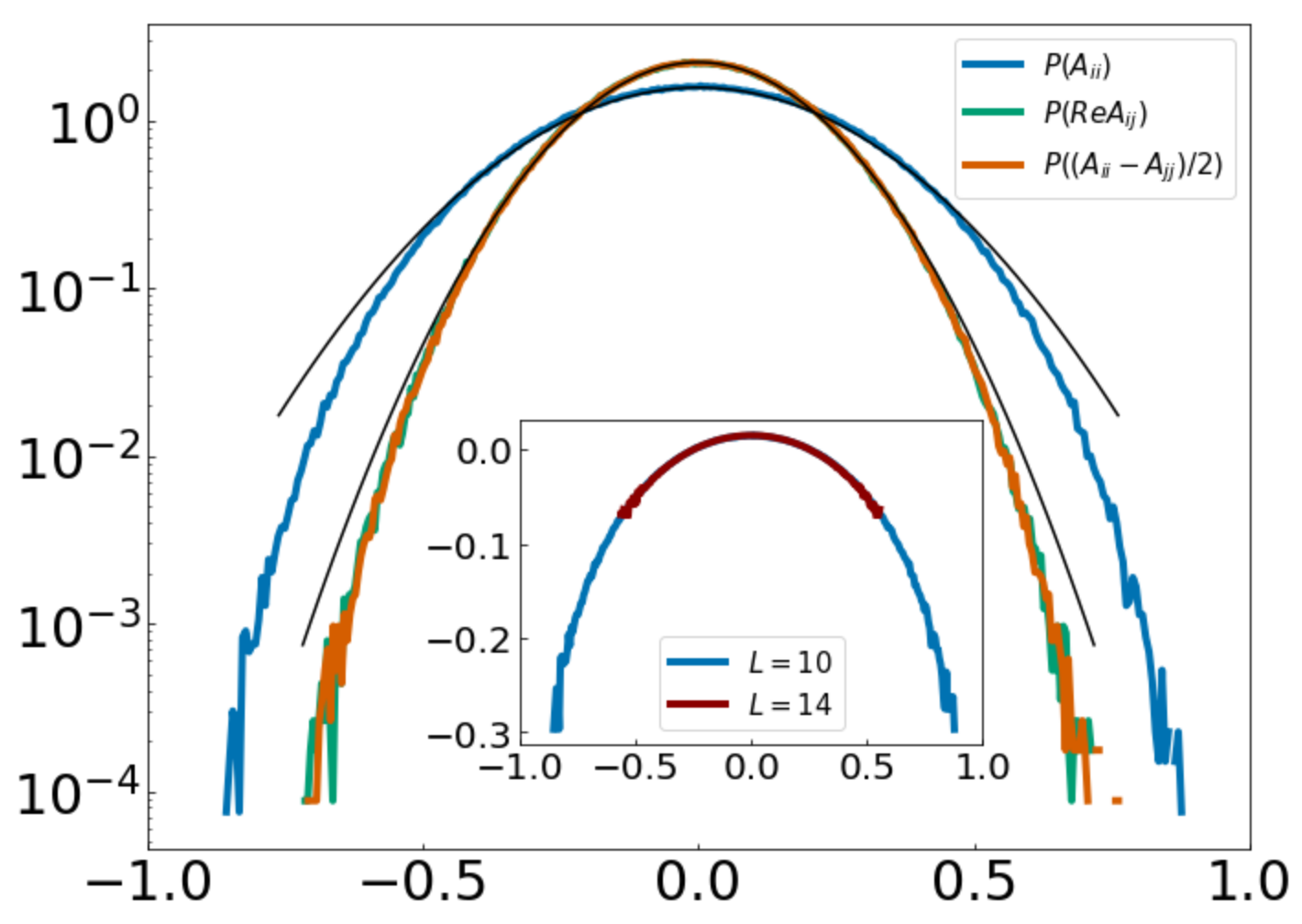}
\caption{SYK model with L=10 Majorana fermions. 
We show $P(A_{ii})$,  $P(\text{Re} A_{ij})$ 
and $P((A_{ii}-A_{jj})/2)$ for the observable $A=i \psi_{\alpha} \psi_{\beta}$ over all pair $\alpha$ $\beta$. 
We choose $i$ in the middle of the spectrum and $j=i+1$.
Comparison with a Gaussian is shown with black lines. 
In the inset we show the collapse of $\log P_d(A_{ii})/N+C$ with $N=2^{L/2}$ and $C$ is an arbitrary constant
for $L=10$ and $L=14$.
}\label{Fig_SYK}
 \end{figure}

~\\
{\bf Applications to quantum systems.} \\

We now check if in ergodic quantum systems the result  we have obtained
relating the distribution of diagonal and off-diagonal matrix elements applies. 
Note that what
was a theorem for a matrix model is here a leap of faith.
For a true matrix model the relation is  valid for any $N$ while for a quantum system one may expect that a
 random matrix ETH regime sets in for $N$  sufficiently large.
 The particular case  of variances of elements in 
a  Gaussian case  was already discussed in Refs. \cite{mondaini2017,hamazaki2019}.

In the following, in the spirit of ETH we look at matrix elements of physical
observables in the energy eigenbasis  (or in the basis of the evolution operator
for Floquet systems) and we order eigenstates
according to their energy or their phase.

As a first example we consider a paradigmatic model of chaos, the Sachdev-Ye-Kitaev (SYK) model~\cite{maldacena2016}.
This is defined considering $N$ Majorana fermions $\{\psi_\alpha,\psi_\beta\}=\delta_{\alpha\beta}$ interacting
via a disordered multibody term:
\begin{equation}
H = \sum_{0\leq \alpha<\beta<\gamma<\delta \leq L} J_{\alpha\beta\gamma\delta} \psi_\alpha \psi_\beta \psi_\gamma \psi_\delta
\end{equation}
where $J_{\alpha\beta\gamma\delta}$ are Gaussian random variables with zero mean and variance 
$\langle J_{\alpha\beta\gamma\delta}^2\rangle = 6/L^3$.
We consider the observable $A = i \psi_{\alpha}\psi_{\beta}$ for all pairs
of $\alpha$ and $\beta$. The Hamiltonian commutes with the parity operator $P=i^{-L/2} \prod_{\alpha=1}^L \psi_\alpha$ and we
restrict ourselves to the sector with eigenvalue $\lambda_{P}=-1$.
In Fig. \ref{Fig_SYK} we show the results for $L=10$ where we compare the distribution
of off-diagonal matrix elements $P(\text{Re} A_{ij})$ with the difference of two diagonal matrix elements $P((A_{ii}-A_{jj})/2)$
in the middle of the spectrum. 
We also show the distribution of diagonal matrix elements $P(A_{ii})$ that is far from 
a Gaussian distribution in the wings, where the
agreement between $P(\text{Re} A_{ij})$ and $P((A_{ii}-A_{jj})/2)$ is still good.
In the inset we show the collapse of the large deviation function $P(A_{ii})$ for two sizes $L=10$ and $L=14$.

\begin{figure}
\centering \includegraphics[angle=0,width=0.4\textwidth]{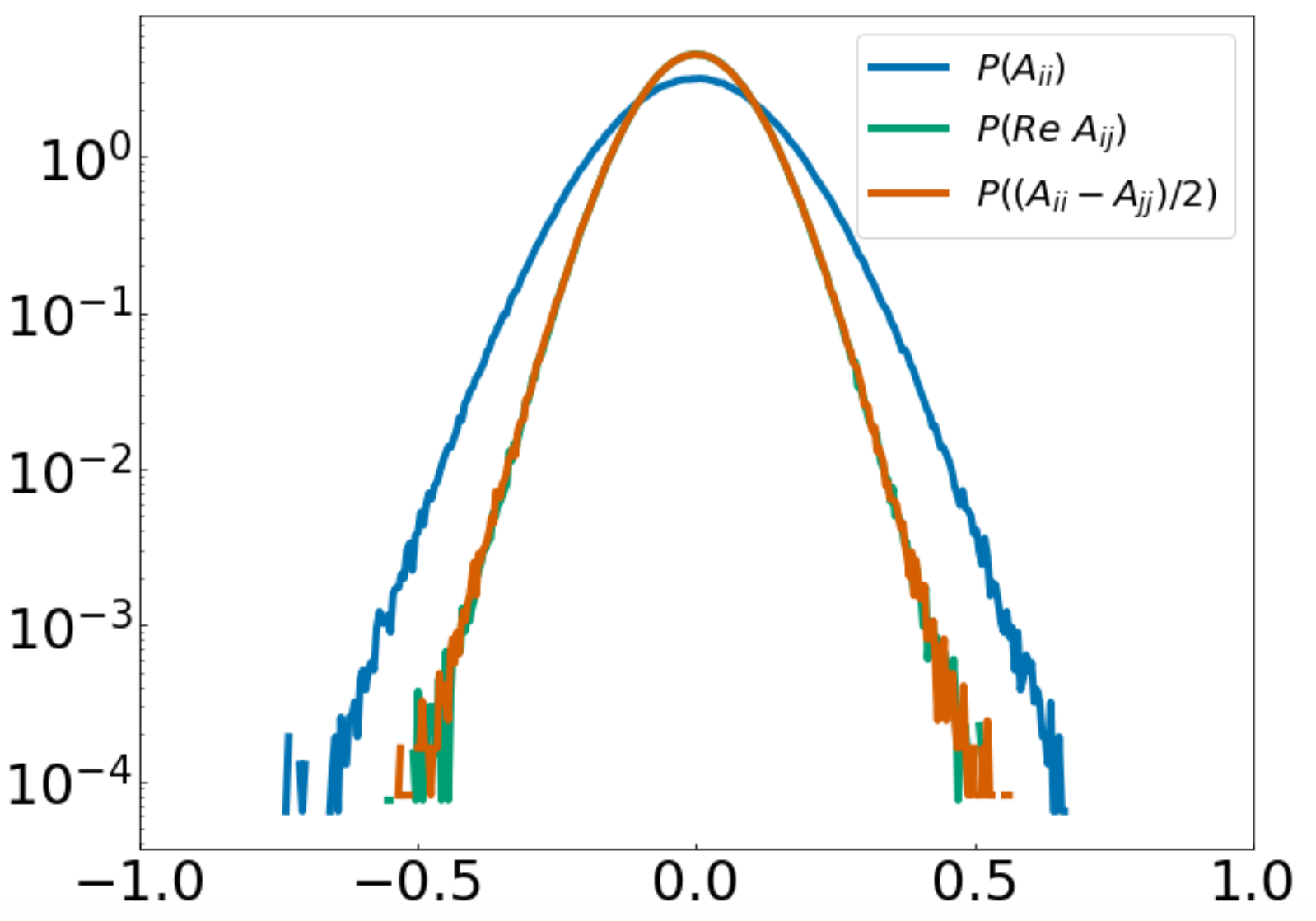}
\caption{Floquet model with $L=6$ sites. We show $P(A_{ii})$,  $P(\text{Re} A_{ij})$ 
and $P((A_{ii}-A_{jj})/2)$ for the observable $\sigma_{L/2}^z$. 
As energy is not conserved we used all eigenstates for each realization
and for each $i$ we choose $j=i+1$.
}\label{Fig_Floquet}
 \end{figure}

Next we consider a Floquet system as the one implemented in \cite{chan2018}.
We take a one dimensional chain and we consider as evolution operator
$U=W_1 W_2$ with
$W_1=U_{1,2} \times U_{3,4} \times \dots \times U_{L-1,L}$ and
$W_2=S^T (U_{1,2} \times U_{3,4} \times \dots \times U_{L-1,L} ) S$ where 
$U_{i,j}$ are random unitary matrix and $S$ is the
shift operator which translate the spins of one site implementing periodic boundary conditions.
We consider as observable $A = \sigma^z_{L/2}$ in the basis of the operator $U$.
In Fig. \ref{Fig_Floquet} we show the result of the distribution of diagonal and off-diagonal
matrix elements for a system of size $L=6$. The deviations from a Gaussian are clear
as well as the agreement between $P(\text{Re} A_{ij})$ and $P((A_{ii}-A_{jj})/2)$.

\begin{figure}
\centering \includegraphics[angle=0,width=0.4\textwidth]{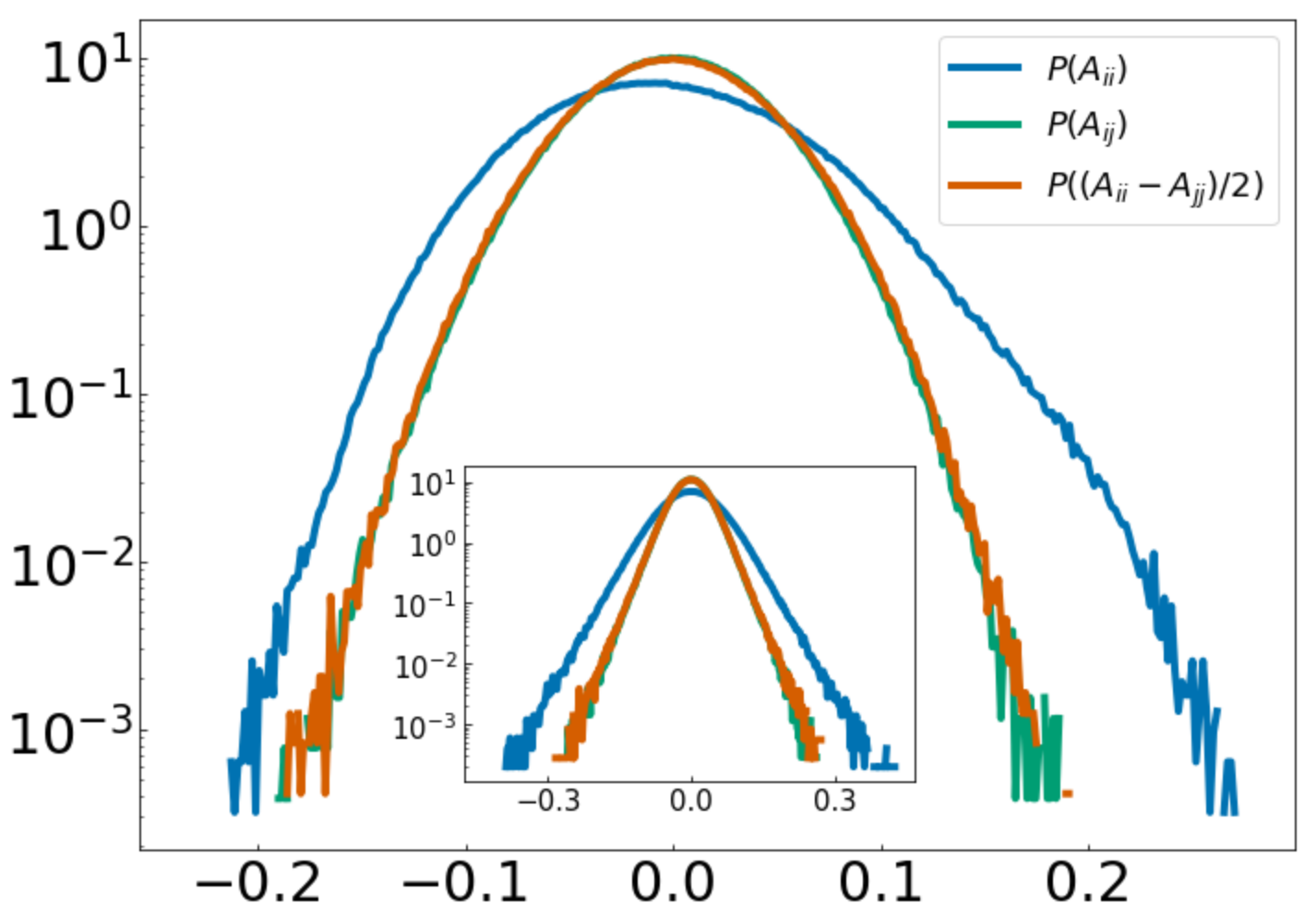}
\caption{Disordered chain with $L=12$ spins at small disorder $h=1$. 
The data correspond to $P(A_{ii})$, $P(A_{ij})$ and $P((A_{ii}-A_{jj})/2)$.
We take $20$ states in the middle of the spectrum and for each $i$ we choose $j=i+1$.
In the main panel we
show the distributions for a given sample of disorder (up to small variations $\delta h_i \in [-0.05,0.05]$)
and one spin, in the inset the distributions averaged over the disorder
and the spins.
}\label{Fig_chain}
 \end{figure}

We finally discuss a disordered spin chain:
\begin{equation}\label{Eq_chain}
H =  \sum_{i=1}^L \left[ J  S_i \cdot S_{i+1} + h_i S^z_i \right]
\end{equation}
 where $S_i$ is a spin $1/2$, the sum is over periodic boundary conditions
 and $h_i$ are random uniform variables between $[-h,h]$.
 In Fig. \ref{Fig_chain} we show the results for $L=12$ and $h=1$
 and we choose $S^z_{L/2}$. 
 In our procedure to construct the histogram we generate a configuration of fields
  and then we vary it by a small amount ($\delta h_i \in [-0.05,0.05]$), to avoid very rare values of disorder fields
  dominating the tails -- an effect enhanced at small sizes.
We obtain a good agreement between $P(A_{ij})$ and $P((A_{ii}-A_{jj})/2)$.
 The deviations from the Gaussian are not very marked for this particular sample but are inevitable given that
 $P(A_{ii})$ is clearly not Gaussian.
 Allowing the fields to vary freely over the disorder gives different distributions
 (as we show in the inset of Fig. \ref{Fig_chain}), and the symmetry of the distributions is clearly recovered.

Let us note that the distributions of off-diagonal matrix elements in many models
 (also considered here) are observed to be Gaussian \cite{haque2017,chan2018b,luitz2016,luitz2017,beugeling2015},
 when relatively small deviations are considered, as is usual in large-deviation theory.
 
 \begin{figure}
\centering \includegraphics[angle=0,width=0.4\textwidth]{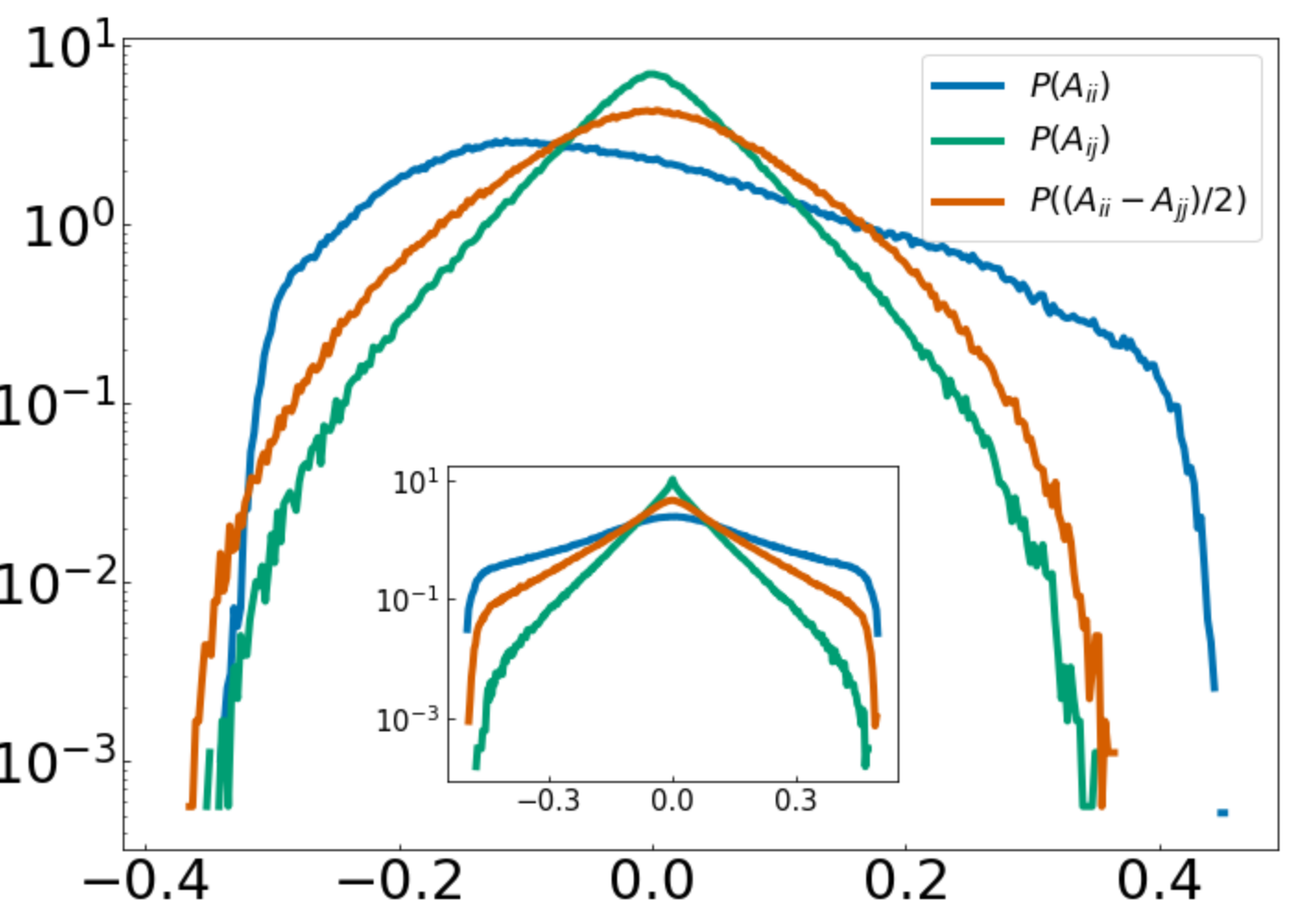}
\caption{Imperfect eigenstate thermalization: disordered chain with $L=12$ spins at larger disorder $h=2$. In the main panel we
show the distributions for a given sample, in the inset the distributions averaged over the disorder
and the spins. As before we take $20$ states in the middle of the spectrum.
}\label{Fig_chain2}
 \end{figure}

 Perhaps the most interesting situation is when the distributions do not coincide.
 This can be seen for instance increasing the disorder in the model (\ref{Eq_chain}).
 Already for $h=2$ and $L=12$ the distributions $P(A_{ij})$ and $P((A_{ii}-A_{jj})/2)$ differ
 as shown in Fig. \ref{Fig_chain2}.
 This value of $h$ is expected to be below the MBL transition \cite{pal2010,luitz2017,vsuntajs2019quantum}, but at these sizes 
 eigenstate thermalization does not hold.
For large  $h$,  in the MBL phase,  the relation is expected to break down for every size.

 {\it Conclusions.} 
 We have introduced a test of eigenstate thermalization inspired by matrix models that can  be used in 
 conjunction with level-statistics  to detect subtle localization properties. 
 The test is a direct check on the eigenvectors of the Hamiltonian rather than an indirect test through its spectrum. \\
 The example of the matrix models
underlines the fact that there is a close connection between the non-gaussian nature of the distribution and
the existence of correlations between matrix elements (which are relevant for the higher-order correlation functions \cite{foini2019}).
 This relation does not apply for level-spacing, which is insensitive the existence or not of correlations between matrix elements.
 
 {\it Acknowledgments} 
 We thank discussions with J. Chalker, A. De Luca, B. Eynard and P. Vivo.
 This work is supported by ``Investissements d'Avenir" LabEx PALM
(ANR-10-LABX-0039-PALM) (EquiDystant project, L. Foini)
 J.K. is supported by the Simons  Foundation Grant No 454943.


%

\end{document}